\documentclass[twocolumn,showpacs,amsmath,amssymb]{revtex4} 
\usepackage{graphicx}
\usepackage{dcolumn}
\usepackage{bm}
\usepackage{braket}
\usepackage{amsmath}
\usepackage{amssymb}
\usepackage[dvips]{color}

\begin{document}

\title{MgB$_{2}$ Nonlinear Properties Investigated Under Localized High RF Magnetic Field Excitation}

\author{Tamin Tai$^{1,2}$}
\author{B. G. Ghamsari $^{2}$}
\author{T. Tan$^{3}$}
\author{C. G.Zhuang$^{3}$}
\author{X. X. Xi$^{3}$}
\author{Steven M. Anlage$^{1,2}$}

\affiliation{$^{1}$Department of Electrical and Computer
Engineering, University of Maryland, College Park, MD 20742-3285,
USA}

\affiliation{$^{2}$Department of Physics, Center for Nanophysics
and Advanced Materials, University of Maryland, College Park, MD  20742-4111, USA}

\affiliation{$^{3}$Physics Department, Temple University,
Philadelphia, PA 19122, USA}

\begin{abstract}
The high transition temperature and low surface resistance of
MgB$_2$ attracts interest in its potential application in
superconducting radio frequency (SRF) accelerating cavities. However,
compared to traditional Nb cavities, the viability of MgB$_2$ at
high RF fields is still open to question. Our approach is to study
the nonlinear electrodynamics of the material under localized RF
magnetic fields. Due to the presence of the small superconducting
gap in the $\pi$ band, the nonlinear response of MgB$_2$ at low
temperature is potentially complicated compared to a single gap
s-wave superconductor (SC) such as Nb. Understanding the mechanisms
of nonlinearity coming from the two-band structure of MgB$_{2}$, as
well as extrinsic sources of nonlinearity, is an urgent requirement.
A localized and strong RF magnetic field, created by a magnetic
write head, is integrated into our nonlinear-Meissner-effect
scanning microwave microscope \cite{T. Tai}.  MgB$_{2}$ films with
thickness 50 nm, fabricated by a hybrid physical-chemical vapor
deposition technique on dielectric substrates, are measured at a
fixed location and show a strongly temperature-dependent third
harmonic response. We propose that several possible mechanisms are
responsible for this nonlinear response.

\end{abstract}


\maketitle

\section{INTRODUCTION}
The discovery of superconductivity in MgB$_{2}$ in January 2001
\cite{J. Nagamatsu} ignited enthusiasm and interest in exploring its
material properties. Several remarkable features, for example a high
transition temperature ($T_{c}\sim $ 40 K ), a high critical field,
and a low RF surface resistance below $T_{c}$, shows great potential
in several applications such as superconducting wires and magnets.
The success of making high quality epitaxial MgB$_{2}$ thin films
provides another promising application as an alternative material
coating on superconducting radio frequency (SRF) cavities \cite{Xi}.
Over the past decade, the accelerating gradient has achieved 59 $MeV/m$
in fine-grain Niobium (Nb) single cell cavity \cite{Geng}. In order to go further, new high
$T_c$ materials with low RF resistance are required for interior
coating of bulk Nb cavities. High quality MgB$_{2}$ thin films may
satisfy the demands for SRF coating materials because such films can
avoid the weak link nonlinearity between grains, and lead to the
possibility of making high-Q cavities \cite{Tajima}.

However, there still exist mechanisms that produce non-ideal
behavior at low temperatures under high RF magnetic fields, such as
vortex nucleation and motion in the film \cite{A. Gurevich}. In
addition, due to the presence of the $\pi$ band and $\sigma$ band,
the intrinsic nonlinear response of MgB$_{2}$ at low temperature is large
compared to single-gap s-wave superconductors \cite{G. Cifariello} \cite{T. Dahm}.
Finally, it has been proposed that MgB$_{2}$ has 6 nodes in its
energy gap \cite{Agassi}. The intermodulation distortion (IMD)
measurements show a strong enhancement at low temperature (T) as
$P_{IMD}(T)$ $\sim$~1/$T^2$ \cite{Oates}, similar to the
characteristics of the d-wave nodal $YBa_{2}Cu_{3}O_{7-\delta}$
(YBCO) superconductor \cite{D. E. Oates1}. Note that the nodal
nonlinear Meissner effect has only been measured by means of
nonlinear microwave techniques up to this point. If MgB$_2$ is a
nodal superconductor, the coating of MgB$_2$ on SRF cavities will
limit the high-field screening response at low temperature and
therefore degrade the performance of the SRF cavities. Based on the
above concerns, the study of MgB$_{2}$ microwave nonlinear response
in the high frequency region (usually several GHz in SRF
applications) can reveal the dissipative and nondisipative nonlinear
mechanisms and perhaps enable application of MgB$_{2}$ films as
cavity coatings.

In our experiment the localized harmonic response of superconductors
is excited by a magnetic write head probe extracted from a
commercial magnetic hard drive \cite{T. Tai}. Based on the gap
geometry of the magnetic write head probe, sub micron resolution is
expected. We present our observation of the nonlinear response of
high quality MgB$_{2}$ films below $T_{c}$. These films were grown
on (0001) sapphire substrates by the hybrid physical-chemical vapor
deposition technique (HPCVD). A (0001)-oriented MgB$_2$ film with
(10$\overline{1}$0) in-plane epitaxial structure was determined by
$\theta$-2$\theta$ and $\phi$ scans in X-ray diffraction,
respectively. A detailed description of the growth technique and
their structural analysis has been reported before \cite{X. Zeng}.

It should be noted that the SRF cavities function at very low
temperature and in RF magnetic fields of varying strength, depending
on location in the cavity. However our microscope functions at
temperatures down to 5 K, and with a localized high RF magnetic
field. Therefore our microscope is best suited for finding the
localized electromagnetic response of the surface. We expect to find
electromagnetic contrast due to surface defects on the materials of
the cavities. In this paper, we report the MgB$_2$ experimental
nonlinearity data from localized areas. These data will be
interpreted as a combination of several nonlinear mechanisms
including intrinsic nonlinear responses \cite{T. Dahm}\cite{Agassi}
and vortex nonlinearity \cite{A. Gurevich}.

\begin{figure}[Experiment-technique]
   \centering
   \includegraphics*[width=3 in, angle=0]{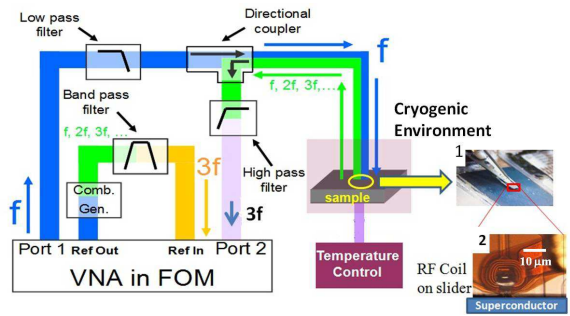}
   \caption{Set up of phase-sensitive harmonic measurement in nonlinear microwave microscopy. The frequency offset mode (FOM) of a vector network analyzer (VNA :model PNA-X N5242A) is used in this measurement.
   Inset 1. shows the magnetic write head probe assembly on top of a superconducting thin film and inset 2. is a picture of the RF micro-coil of the probe above the sample.}
   \label{figure_setup}
\end{figure}

\section{EXPERIMENTAL SETUP}
The experimental setup for amplitude and phase measurements of the
superconductor harmonic response is shown in Fig.
\ref{figure_setup}. An excited wave (fundamental signal) at
frequency $f$ comes from the vector network analyzer (VNA) and is
low-pass filtered to eliminate higher harmonics of the source
signal. This fundamental tone is sent to the magnetic write head
probe to generate a localized RF magnetic field on the
superconductor sample. Two insets in Fig. \ref{figure_setup} show
close-up views of our magnetic write head probe on superconducting
samples.  Due to the intense nature of this field, the
superconductor responds by generating screening currents at both the
fundamental frequency and at harmonics of this frequency.  The
generated and coupled harmonic signal is high-pass filtered to remove the
fundamental signal $V_f$ and an un-ratioed measurement of $V_{3f}$
is performed on port 2 of the VNA. In order to get a phase-sensitive
measurement of the $3^{rd}$ harmonic signal coming from the
superconducting sample, a harmonic generation circuit is connected
to provide a reference $3^{rd}$ harmonic signal, and the relative
phase difference between the main circuit and reference circuit is
measured. Further detail about this phase-sensitive measurement
technique can be found in Ref. \cite{D. Mircea}. In this way we
measure the complex third harmonic voltage of $V_{3f}^{sample}(T)$
or the corresponding scalar power $P_{3f}^{sample}(T)$. The lowest
noise floor in our VNA is -130 dBm for the un-ratioed power
measurement. A ratioed measurement of the complex
$V_{3f}^{sample}(T)/V_{3f}^{ref}$ is also performed at the same
time. An alternative method to lower the noise floor is to remove
the VNA and use a stable synthesizer on port 1 and a phase-locked spectrum
analyzer on port 2. The noise floor of our spectrum analyzer (model
$\sharp$: ESA-E E4407B) is -145 dBm. In this paper we only discuss
the unratioed measurements of $P_{3f}^{sample}(T)$ and qualitatively
discuss the mechanisms of third harmonic response of the MgB$_2$
film.

\section{THIRD ORDER NONLINEAR MEASUREMENT RESULTS}
\begin{figure}[tb]
    \centering
    \includegraphics*[width=2.7in]{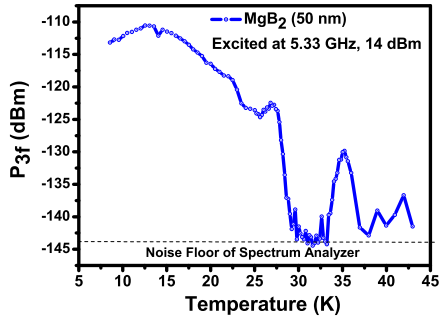}
    \caption{Temperature dependence of 3rd harmonic power $P_{3f}$ from a 50 nm thick MgB$_{2}$
    of group A measured with an excited frequency of 5.33 GHz at +14 dBm.}
    \label{MgB2_50nm_A}
\end{figure}

The measurement of the $3^{rd}$ order harmonic power ($P_{3f}$) is
performed near the center of several epitaxial MgB$_{2}$ films with
the same thickness, 50 nm. The $T_c$ of these samples are around 32
K$\sim$35 K measured by the four point resistance method. These
samples are all grown on sapphire substrates under the same
deposition conditions by the HPCVD method. These samples can be
grouped into two classes: Group A are the samples which are not well
isolated from the ambient environment after growth. Group B are the
samples which are kept in desiccated conditions immediately after
deposition. At least two samples are measured from each group to examine their RF microwave properties.
Fig. \ref{MgB2_50nm_A} shows a representative
temperature dependent $P_{3f}(T)$ curve for the sample from group A
at the excited frequency 5.33 GHz and excited power +14 dBm. Above
40 K a very small signal begins to arise above the noise floor of
the spectrum analyzer. This $P_{3f}$ is from the magnetic write head
probe itself. We have measured the $P_{3f}$ of the magnetic probe when it is placed on
the surface of a bare sapphire substrate and in general this probe
nonlinearity is negligible at excited powers under +14 dBm. Although
excited powers above +14 dBm excites stronger nonlinearity from the
probe, this nonlinearity is almost temperature independent in the
Helium cooling temperature range. Therefore probe nonlinearity, if
it is present, can be treated as a constant background signal above
the noise floor of the spectrum analyzer/VNA. The mechanism of probe
nonlinearity is the hysteretic behavior of the yoke material,
and has been discussed previously \cite{T. Tai}.

From Fig. \ref{MgB2_50nm_A}, a clear $P_{3f}(T)$ peak centered at 35
K shows up above the noise floor. This peak arises from the
modulation of superconducting order parameter near $T_c$ due to the
enhanced sensitivity of superconducting properties as the superfluid
density decreases to near-zero levels. This peak at $T_{c}$ is also
phenomenologically predicted by Ginzburg-Landau theory, and is
discussed further below.

We also note the onset of a temperature dependent $P_{3f}$
nonlinearity below 29 K, followed by a peak near 27 K, and then a
gradually increasing $P_{3f}$ down to 12.5 K. Finally, the $P_{3f}$
decreases below 12.5 K.

\begin{figure}
    \includegraphics*[width=2.9 in]{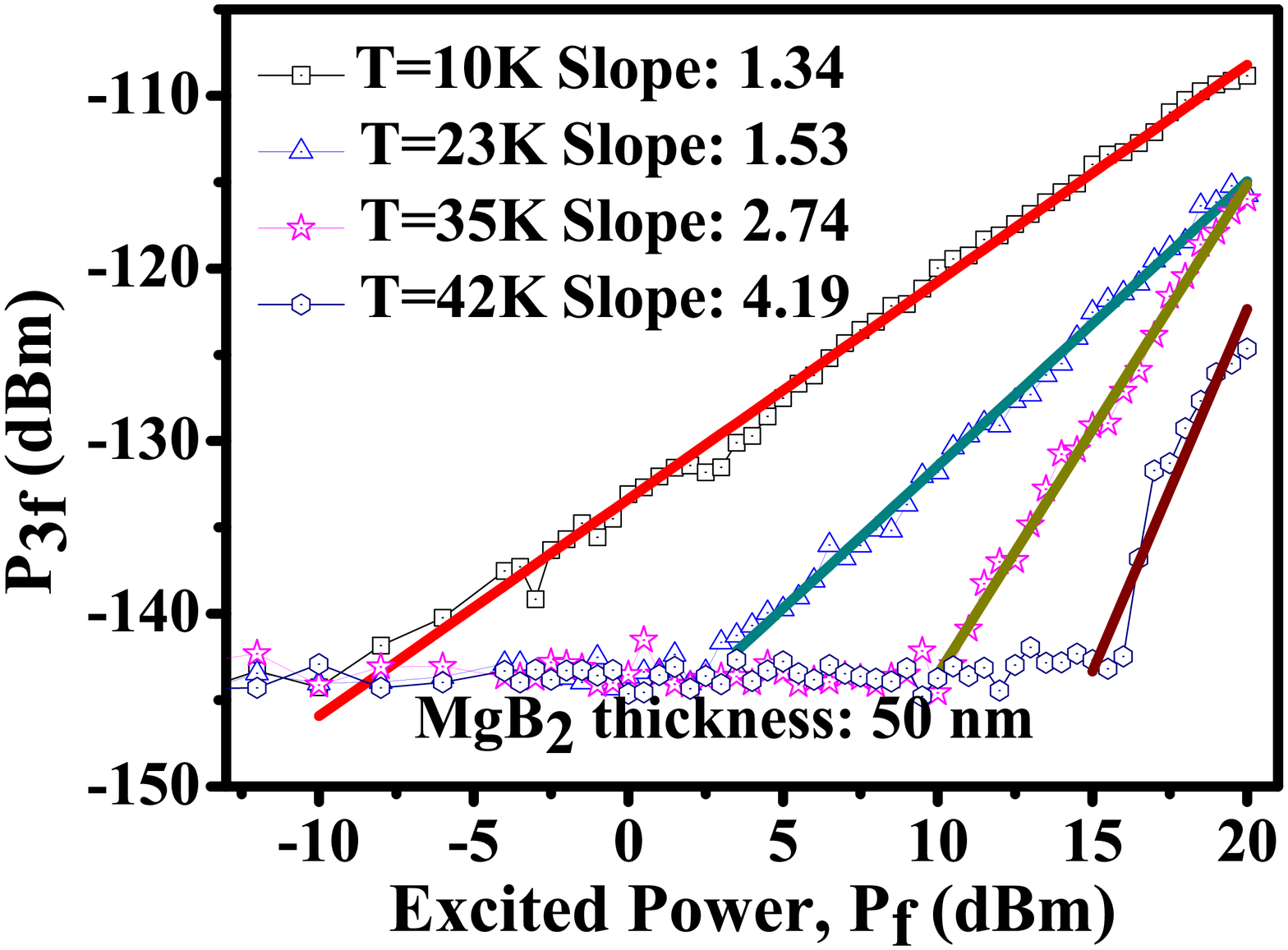}
    \centering
    \includegraphics*[width=2.9 in]{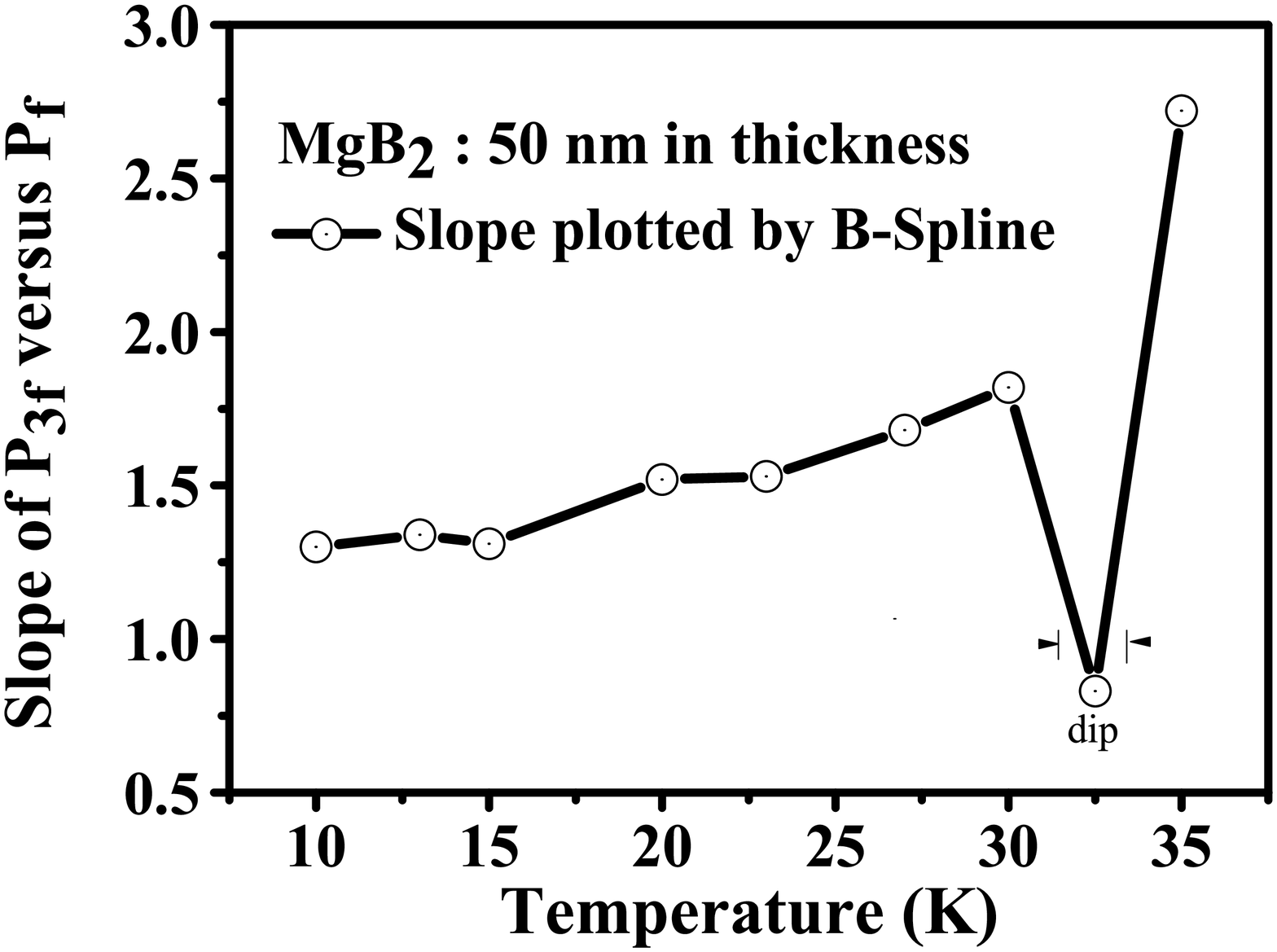}
    \caption{(top frame) Power dependence of $P_{3f}$ on $P_{f}$ for the
    50 nm thick MgB$_{2}$ film of group A. (bottom frame) Fitted slope at selected
    temperatures for the film of group A. The marked dip describes an
    almost nonlinearity-free region from the MgB$_2$ sample and its
    small slope is likely due to the probe nonlinearity.}
    \label{MgB2SlopeAll}
\end{figure}

Measurements of the dependence of $P_{3f}$ on $P_{f}$ are shown in
Fig. \ref{MgB2SlopeAll} for the 50 nm thick MgB$_{2}$ film (group A)
at some selected temperatures. In the normal state of MgB$_{2}$
(T=42K), the measured nonlinearity comes from the probe itself and
shows a slope steeper than 3 at high excited power above +15 dBm.
Near $T_c$, the slope is 2.74, close to the value of 3 as predicted for the
intrinsic nonlinear response \cite{John Lee}. Based in part on this evidence, we
believe that in the high temperature region close to $T_{c}$, most of the
$P_{3f}$ comes from the intrinsic nonlinear mechanism related to the modulation of the order parameter near $T_c$. In the intermediate and low
temperature regime, the slopes of $P_{3f}$ vs. $P_{f}$ are between
1$\sim$2. This value is similar to that predicted by many
phenomenological models (between $1 \sim 2$) for YBCO films at low temperature \cite{D. E. Oates1} \cite{D. Agassi}, implying that several possible nonlinear mechanisms are involved at low temperature for MgB$_2$ films. It should be noted that the low temperature
nonlinearity can be easily excited at low power. Fig.
\ref{MgB2SlopeAll} shows the $P_{3f}$-$P_{f}$ slope evolution from an
intrinsic nonlinear region around $T_c$ to a regime with different behavior at
lower temperature.

\begin{figure}
\centering
\includegraphics*[width=3in]{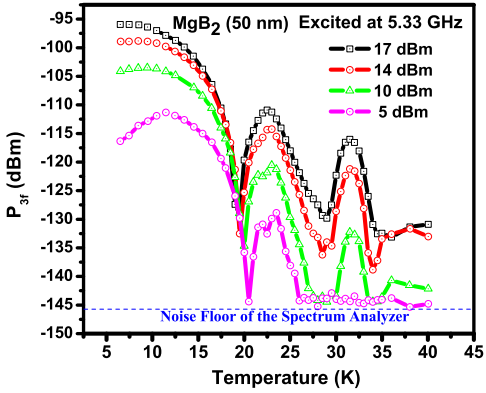}
\caption{Temperature dependence of 3rd harmonic power $P_{3f}$ from
a 50 nm thick MgB$_{2}$ of group B measured with an excited
frequency of 5.33 GHz at different excited powers.}
\label{MgB2_50nm_B}
\end{figure}

The representative curve of $P_{3f}(T)$ from a sample of group B is
shown in Fig. \ref{MgB2_50nm_B}. Compared with the measurement of
$P_{3f}(T)$ from the sample of group A, as shown in Fig.
\ref{MgB2_50nm_A}, many temperature dependent nonlinear features are
consistent and reproducible.  From Fig. \ref{MgB2_50nm_B}, the first
peak at 32 K represents the first $T_c$ of this sample. A second
peak at 22.5 K, similar to that at 27 K (Fig. \ref{MgB2_50nm_A}) is
distinctly visible in Fig. \ref{MgB2_50nm_B}. The only difference is
a second deep dip around 20 K for the sample from group B, versus a
shallow dip at 25 K for the sample from group A in the $P_{3f}(T)$
measurement. The position of the dip and its depth also change with
the excited power. This sharp dip indicates the near cancelation of
all nonlinear mechanisms at this temperature. For both groups of
samples, the nonlinearity gradually increases with reduced
temperature below the second dip, followed by a saturation and
finally a decrease at temperatures below 10 K, at least for lower
excitation power.

\begin{figure}
\centering
\includegraphics*[width=3 in]{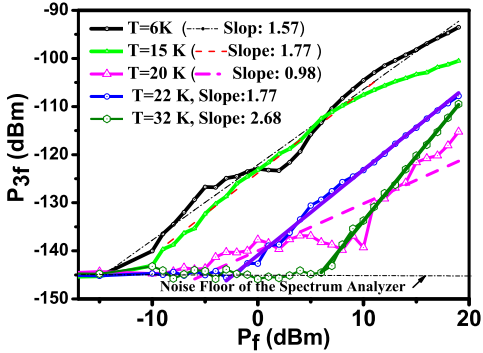}
\caption{Power dependence of $P_{3f}$ on $P_{f}$ for the 50 nm thick
MgB$_{2}$ film of group B.} \label{MgB2_slope_B}
\end{figure}

Fig. \ref{MgB2_slope_B} shows the dependence of $P_{3f}$ on $P_{f}$
for the sample from group B. At $T_c$ (32 K), the slope of the power
dependence is almost 3. Below $T_c$, the slope drops to a value
between 1$\sim$2, the same as many published results on MgB$_2$
\cite{D. E. Oates2} \cite{Gallitto}, excluding the points near 20 K
at which the slope drops below one. This point (the second dip)
shows very small nonlinearity from the superconductor below 10 dBm
excitation. For high excitation power (above 10 dBm), the majority
of the nonlinearity signal comes from the probe itself. In addition,
at some temperatures (20 K, 15 K, 6 K), the power dependence curve
shows several changes of slope. This implies that the nonlinear
behavior is more complex at intermediate and low temperatures.
Therefore based on the similar nonlinear behavior of many measured
MgB$_2$ samples, we next discuss several possible mechanisms that
may account for the basic common features of our experimental data.

\section{Intrinsic Nonlinearity of MgB$_2$ around $T_c$}
\begin{figure}
\centering
\includegraphics*[width=2 in]{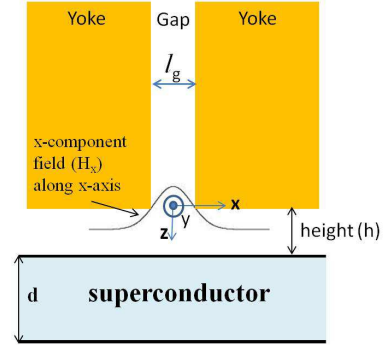}
\caption{Schematic illustration of the 2D magnetic write head on the
top of the superconductor. The origin is at the bottom of the head
centered at the middle of the gap. Gap length ($l_g$) and film thickness (d) are not to scale.} \label{Karlqvist treatment}
\end{figure}

The modulation of superconducting order parameter generates
nonlinearity under excitation by an applied field. While the
excitation field is much lower than the critical field, nonlinearity
will be generated due to the perturbation of the ordered parameter. This phenomenon is
generally known as a nonlinear Meissner effect (NLME). When the
excitation field approaches the critical field, the mechanism of
intrinsic nonlinearity is still similar to the NLME and becomes more significant. For example, around $T_c$, this nonlinearity comes from the backflow of excited quasiparticles
in a current-carrying superconductor, which results in an effective
decrease of the superfluid density \cite{T.
Dahm}\cite{Agassi}\cite{Nicol}. A two band quasiparticle backflow calculation has
been applied to the MgB$_{2}$ intrinsic nonlinearity. Based on
the work of Dahm and Scalapino\cite{T. Dahm}, the temperature and
induced current density dependent superfluid density $n_{s}(T,J)$
can be written as
\begin{equation}\label{eq:superfluid}
    \frac{n_{s}(T,J)}{n_{s}(T,0)}=1-({J \over J_{NL}})^2 ;
    J_{NL}={J_{c,\pi} \over \sqrt{b_{\pi}(T)+b_{\sigma}(T){J_{c,\pi}^{2}
\over J_{c,\sigma}^{2}} }}
\end{equation}
where b$_{\sigma}$ and b$_{\pi}$ are the temperature dependent
nonlinear coefficients for the $\sigma$ band and $\pi$ band,
respectively, and their values are defined in reference \cite{T.
Dahm}. Here $J_{c,\sigma}=4.87\times10^8 A/cm^2$ and
$J_{c,\pi}=3.32\times10^8 A/cm^2$ are the pair-breaking current
densities for the two bands. For a 50 nm thick MgB$_2$ thin film,
the generated third harmonic power $P_{3f}(T)$ is estimated by
substituting $J_{NL}$ into the following equation derived from the
assumption of $d$ (thickness)$<<\lambda$ (penetration depth)
\cite{John Lee}
\begin{equation}\label{eq:P3f}
\begin{split}
    P_{3f}(T)=\frac{\omega^2\mu_{0}^2\lambda^4(T)\Gamma^2(K_{RF})}{32Z_
    {0}d^{6}J^4_{NL}(T)} \propto P_f^3 ;\\
    \Gamma(K_{RF})=I_{tot}\int\frac{\int K^4(x,y)dx}{(\int K_{y}dx)^2} dy
\end{split}
\end{equation}
where $\omega$ is the angular frequency of the incident wave,
$\lambda (T)$ is the temperature dependent magnetic penetration
depth, $Z_0$ is the characteristic impedance of the transmission
line in the microscope, ${I_{tot}}$ is the total current flowing
through a crossection right beneath the bottom of the probe, and
$\Gamma(K_{RF})$ is a geometry factor. The value of geometry factor
depends on the distribution of surface current density ($K_{RF}(x,y))$
on the superconducting plane.
We calculate the surface current from the Karlqvist equation \cite{S. X. Wang}, which gives
the magnetic field distribution outside the gap in the x-z plane as schematically shown in Fig. \ref{Karlqvist treatment}. Under the treatment of the Karlqvist 2D assumption, the fields are invariant in
the y-direction along the gap. By only considering the x-component
of magnetic field ($H_x$) on the superconducting surface, which is doubled
with respect to free space by the boundary condition of the surface, the surface current $K_y$ can be written as
\begin{equation}\label{Karlqvist}
K_y=\frac{2B_g}{\mu_0}\arctan{[\frac{l_g z}{x^2+z^2-(l_g/2)^2}]}
\end{equation}

\begin{figure}
\centering
\includegraphics*[width=3 in]{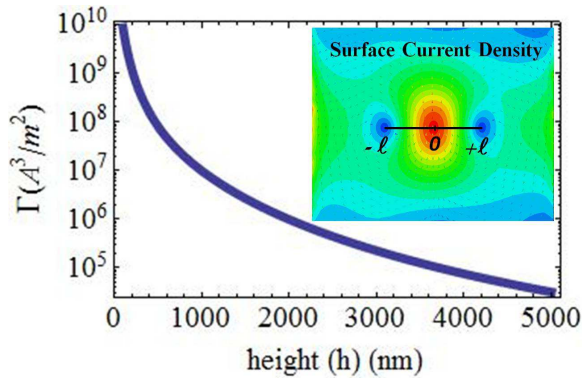}
\caption{Calculation of height dependence (h is the +Z direction) of the probe geometry factor $\Gamma(K_{RF})$ for the longitudinal magnetic write head probe. The black line in the inset shows the integral path of Eq. (\ref{Gamma}) on the plot of the surface current density simulated by the ANSYS High Frequency Structure Simulator
(HFSS) under the assumption that the magnetic writer is 2 $\mu m$ away from a perfect conductor surface. Red corresponds to large current while blue corresponds to small current.
The value of $\ell$  is on the scale of the outside edge dimension of the magnetic yoke, which is 1.5 $\mu m$  in the HFSS simulation, and is effectively infinite in the Karlqvist calculation since the return currents are not considered.} \label{Height_Dependence}
\end{figure}

where $B_g$ is the maximum field strength inside the gap, $l_g$ is the length of the gap and x, z are distances in the x and z directions with the origin centered in the middle on the bottom of the gap. In reality, the gap is of finite thickness in the y direction as
shown in the schematic yoke picture of Fig. \ref{yoke} with
thickness $l_y$. By assuming that the currents are uniform in the
y-direction between $y=0$ and $y=l_y$ and ignoring the return
currents, the current distribution geometry factor $\Gamma$ can be approximated as
\begin{equation}\label{Gamma}
\Gamma=\frac{l_y \int_{-\ell}^{\ell}{K_y^4}dx}{I_{tot}};\qquad
I_{tot}=\int_{-\ell}^{\ell}{K_y}dx
\end{equation}
where $\ell$ is the distance from the center of maximum current to the node of minimum current as shown in the inset of Fig. \ref{Height_Dependence}. In addition, Fig. \ref{Height_Dependence} shows the height ($h$) dependence of
$\Gamma$ when evaluating $K_y$ at $z=h$. In our calculation, we assume $B_g=1$ $Tesla$, $l_g=100$ $nm$ and  probe height $h$=2 $\mu m$ above the superconducting surface. Therefore the value of
the $\Gamma$ can be estimated to be $8.3*10^5$ $A^3/m^2$. We note that the probe geometry factor is a strong function of height and can lead to surface fields above the critical field of the superconductor \cite{John Lee2}.

\begin{figure}[tb]
    \includegraphics*[width=3in]{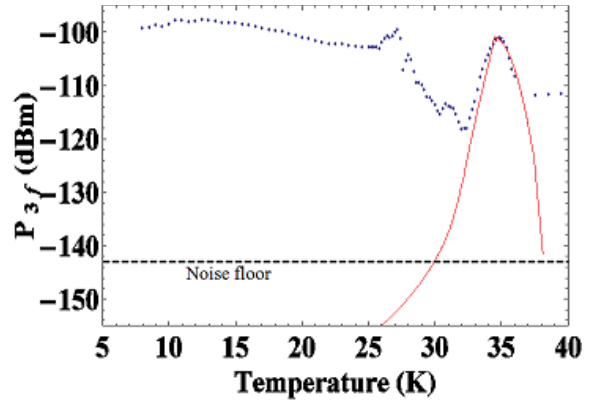}
    \caption{The blue dots are the MgB$_2$ $P_{3f}(T)$ data (from group A) at an excitation frequency of 5.33 GHz and excitation power
    of 18 dBm. The solid red line is the simulated result of the intrinsic nonlinearity from superconducting order parameter modulation
    near $T_c$ of MgB$_{2}$ with thickness 50 nm under the assumption that the magnetic probe provides a field described by a
    geometry factor $\Gamma= 8.3*10^{5}$ $A^{3}/m^{2}$. Other parameters used in this calculation include $\lambda(0 K)$= 100
    $nm$, $J_{NL}(T\rightarrow 0 K)\simeq 6.5*10^{12} A/m^2$, $\lambda_{cutoff}$= 800 $nm$, $J_{cutoff}= 4.2*10^{11}$ $A/m^2$ and $T_c$= 34.6 $K $
    with a standard deviation of Gaussian spread of $\delta T_c$= 1.3 $K$.
    The noise floor in the experiment is -143 dBm.}
    \label{MgB2NLME}
\end{figure}

Finally, the $P_{3f}(T)$ calculated results from (\ref{eq:P3f}) for
the 50 nm thick film at a 5.33 GHz excited frequency is shown as the
solid red line in Fig. \ref{MgB2NLME}, assuming the cutoff of
$\lambda (T\rightarrow T_c), J_{NL}(T\rightarrow T_c)$ and a
Gaussian distribution of $T_c$ \cite{John Lee}. This intrinsic
response has measurable values above the noise floor only in the
high temperature region near $T_{c}$. The experimental data of the
MgB$_{2}$ films from group A under a +18 dBm, 5.33 GHz microwave
excitation is shown as blue dots. It is clear that this model
predicts very low nonlinear response at low temperature. Hence other
mechanisms must be responsible for $P_{3f}(T)$ at temperatures below
$T_c$.
\section{Nonlinearity Due to Second $T_c$ with Leggett Mode at Low Temperature}
\begin{figure}
    \includegraphics*[width=3.4in]{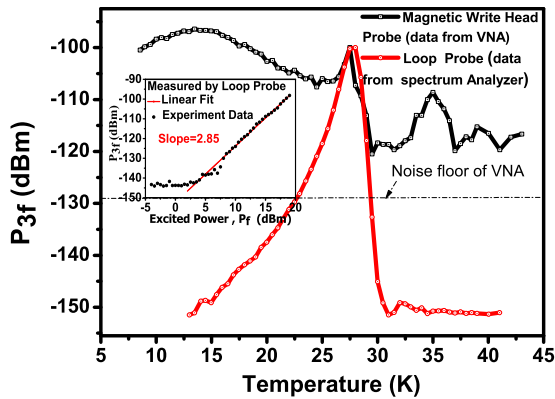}
    \caption{Temperature dependence of $P_{3f}$ from a 50 nm thick
    MgB$_{2}$ film (group A) measured by the loop probe (red curve), and magnetic write
    head probe (black curve), respectively. The excited frequency for both
    measurement is 5.33 GHz. The excited power is 20 dBm and 14 dBm for
    loop probe and magnetic write head probe, respectively. Note that the
    VNA is used to perform the measurement with the write head probe
    and the spectrum analyzer is used in the measurement done by the loop
    probe. The VNA has higher noise floor $\sim$ -130 dBm. The inset shows
    the $P_{3f}$ vs $P_{f}$ dependence for the loop probe measurement at 27.8 K.}
    \label{MgB2loopcompare}
\end{figure}

Theoretically, the nonlinear response of a two-band superconductor
should show a strong peak at the second $T_c$ for completely
decoupled bands \cite{Nicol}. With increasing interband coupling,
the peak due to the second $T_c$ will gradually shift to higher
temperature and have a reduced peak value \cite{Nicol}. Therefore
based on theory, it is possible that in addition to the nonlinearity
coming from the first $T_c$, a proximity enhanced second $T_c$ also
contributes to the intrinsic nonlinearity. From the $P_{3f}(T)$ data
shown in Fig. \ref{MgB2_50nm_A} and Fig. \ref{MgB2_50nm_B}, the
second peak at 27 K (Fig. \ref{MgB2_50nm_A}) or at 23 K (Fig.
\ref{MgB2_50nm_B}) may be due to this intrinsic mechanism. Another
$P_{3f}(T)$ experiment was carried out with a loop probe which
utilizes a smaller RF  magnetic field and a large excitation area.
The loop probe, providing an almost 1 mT in-plane magnetic field on
the MgB$_2$ surface, is made of a nonmagnetic coaxial cable with its
inner conductor (200 $\mu m$ in diameter) forming a $\sim$500 $\mu
m$ outer-diameter semicircular loop shorted with the outer conductor
\cite{John Lee2}. This measurement is performed on the MgB$_2$
sample from group A and the loop probe is positioned on the same
region of the sample where the magnetic write head probe was placed.
In order to compare to the result measured by the magnetic write
head probe, both $P_{3f}(T)$ curves are lined up to -100 dBm at
their peaks around 27 K, as shown in Fig. \ref{MgB2loopcompare}. The
loop probe measurement shows only one peak at 27.8 K almost the same
temperature as the second peak measured by the magnetic write head
probe. The lack of a peak at the first $T_c$ for the loop probe
measurement is due to the weak magnetic field and therefore a small
value of $\Gamma$ in Eq. \ref{eq:P3f} at the highest $T_c$. The
inset of Fig. \ref{MgB2loopcompare} shows the power dependence of
$P_{3f}$ on $P_{f}$ measured by the loop probe at the peak
temperature. The slope of $P_{3f}$ on $P_{f}$ is 2.85, very close to
3, the theoretical value for the intrinsic nonlinear Meissner
effect. However, the slope of $P_{3f}$ on $P_{f}$ obtained with the
magnetic write head probe at this temperature is just 1.68. This implies that our
magnetic write head probe excites another nonlinear mechanism in
this temperature region and there is interference with the
nonlinearity from the second $T_c$. The magnetic write
head probe provides more intense and localized parallel magnetic fields
on the superconductor sample surface. A comparison of magnetic fields
generated by the magnetic write head probe and the loop probe are
reported in Ref. \cite{T. Tai}. It is unclear why the nonlinearity
from the proximity-enhanced second $T_c$ is more significant than
that from the first $T_c$ in the loop probe measurement.

Besides the intrinsic nonlinearity from the proximity-enhanced
second $T_c$, an additional intrinsic nonlinearity arising from
Josephson coupling between the $\sigma$ and $\pi$ bands would be
expected \cite{Gurevich3} below the second peak at 27 K in Fig.
\ref{MgB2_50nm_A} or at 23 K in Fig. \ref{MgB2_50nm_B}. This
nonlinear response (arising from the Leggett mode) comes
from the variable phase difference of the superconducting order
parameters in the two bands of MgB$_2$, and will only be excited
when a nonequilibrium charge imbalance appears both at short length
scales and at temperatures where the proximity-induced $\pi$ band
becomes superconducting \cite{Gurevich3}. In our magnetic write head
experiment, the perpendicular component of the RF magnetic field
results in a charge imbalance and excitation of the Leggett
mode would be expected. From the experimental data of $P_{3f}(T)$ in
Fig. \ref{MgB2_50nm_A} and Fig. \ref{MgB2_50nm_B}, the nonlinearity
below the second peak gradually increases with decreasing
temperature before saturation. In this temperature regime, the
nonlinearity could arise from the Leggett mode mechanism, although
no calculation of this nonlinearity exists, to our knowledge.
However the observed temperature dependence of $P_{3f}(T)$ is
reminiscent of that arising from Josephson weak links \cite{Jeffries} or Josephson vortices
in a large Josephson junction \cite{John Lee2}. The absence of this signal in the macroscopic loop probe measurement is consistent with the charge imbalance mechanism.\\
\section{Nonlinearity From the Reported nodal gap symmetry}
Although MgB$_2$ is commonly believed to be a conventional s-wave
superconductor, Agassi, Oates and Moeckly claim that MgB$_2$ has
line nodes in the superconducting gap, and they claim further that
the gap has 6 nodes
as,$\Delta$($\phi$,T)=$\Delta_0$(T)$\ast$sin(6$\phi$) where $\phi$
is the azimuthal angle in the $\widehat{ab}$ plane of the hexagonal
crystal, and $\Delta_0$(T) is the weakly temperature dependent
amplitude of the gap function at low temperatures \cite{Agassi}.
From their IMD measurement on MgB$_2$ films, the temperature
dependent $P_{IMD}(T)$ shows an upturn around $T<10 K$ and increases
as $1/T^2$ \cite{Oates}. Therefore, based on these observations and
proposals, we would also expect our measurement of $P_{3f}(T)$ to
show an increase in the same temperature range. However all of our
experimental data show that $P_{3f}(T)$ tends to decrease at
temperatures $T< 13 K$. If the prediction of the nodal gap symmetry
is correct, the observed downturn may be due to the interference
between the Leggett mode nonlinearity (or some other nonlinearity)
and the nonlinearity from this nodal gap behavior. Another
possibility may be that the RF magnetic fields employed in our experiment are too strong, or the temperatures are not sufficient small to see the intrinsic NLME due to the nodes.
Yet another possibility is the nonlinear response of Andreev bound state, arising from a sign change of the superconducting gap, on the surface of MgB$_2$
\cite{Zare} \cite{Zhuravel}. Further investigation at lower temperature is required.
\section{Nonlinearity From Moving Vortices}
\begin{figure}
    \centering
    \includegraphics*[width=2.3in]{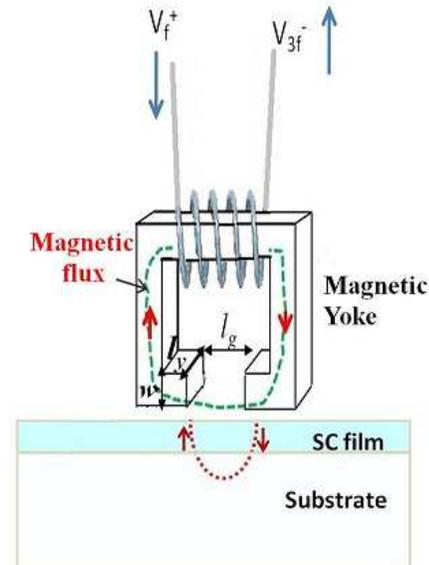}
    \caption{Schematic illustration of the magnetic flux coming from the yoke of the magnetic write head to the superconductor (SC) where $l_y,w$ and $l_g$ represent the width, the thickness, and the length of the gap, respectively.  The length $l_g$ is on the order of 100 nm for our write head probe. The width ($l_y$) is 200 nm and thickness $(w)$ is around 1 $\mu m$. Figure not to scale.}
    \label{yoke}
\end{figure}

Vortex nucleation and penetration into the film induces a dynamic
instability and generates harmonic response \cite{T B Samoilova}.
Considering the relation of the penetration depth of MgB$_2$
($\sim$140 nm) and our film thickness (50 nm), the tendency to
create a straight vortex parallel to the film surface will be
suppressed. However due to the magnetic field distribution from the
magnetic write-head probe, a significant vertical magnetic field is
expected. Fig. \ref{yoke} shows a schematic illustration of our
experiment in which the RF magnetic field from the magnetic write
head probe interacts with the superconductor underneath the probe. A
vortex and an antivortex nucleate perpendicular to the film and will
move under the influence of the RF screening currents. One can model this situation with an
equivalent point magnetic dipole that is placed above the
superconducting thin film \cite{Gilson}. The creation, motion and
destruction of perpendicular vortex and antivortex pairs will
generate high order harmonic response in the experiment. The
nonlinear measurements in films from group A and group
B would have nonlinearity from moving vortices in the entire temperature region
under high RF magnetic field. In addition, vortex nonlinearity due to weak link coupling between each grain under the localized RF field may be another
mechanism. The nonlinearity from weak link vortices in a YBCO bi-crystal grain
boundary has demonstrated a significant increasing nonlinearity with decreasing temperature following the temperature dependence of the critical current of the junction \cite{John Lee2}. From the $P_{3f}(T)$
measurement of MgB$_2$ in Fig. \ref{MgB2_50nm_A} and Fig. \ref{MgB2_50nm_B}, the trends of temperature dependent nonlinearity below 30 K are very similar to the nonlinearity from the
weak link vortices in a YBCO granular structure, except for the deep dip around 20 K in Fig. \ref{MgB2_50nm_B}. Comparing the films of group A to that of group B, both Abrikosov
vortex nonlinearity and weak link vortex nonlinearity of the films
from group A would be expected to be more significant due to the
exposure to air, which will degrade the film and therefore
decrease the lower critical field of the weak links and grains. Models based on the
creation and annihilation of perpendicular vortices and the weak
link vortices are currently under development.

\section{CONCLUSIONS}
A strongly temperature-dependent third harmonic response is found
from near field microwave microscopy measurement on high quality
MgB$_2$ films. From the investigation of the third harmonic response
as a function of temperature and input power level, the nonlinear
mechanisms in high quality MgB$_2$ films appear to be quite complex.
The nonlinear response near $T_c$ can be well understood by a model
relating modulation of the superconducting order parameter near $T_c$.
However the nonlinear response at temperature less than $T_c$ shows
several different possible nonlinear mechanisms. The first is the
intrinsic nonlinearity from the proximity-induced second $T_c$. The
second is the intrinsic nonlinearity arising from Josephson coupling
between the $\sigma$ and $\pi$ bands. The third is the potential
nonlinearity from the proposed nodal gap symmetry of MgB$_2$.
Finally is the nonlinearity due to the perpendicular vortex pairs as
well as the inevitable weak link vortices created in the high
quality MgB$_2$ films by the high RF field probe.
\section{Acknowledgement}
This work is supported by the US Department of Energy/ High Energy
Physics through grant $\#$ DESC0004950, and also by the ONR AppEl,
Task D10, (Award No.\ N000140911190), and CNAM. The work at Temple University is supported by DOE under grant No. DE-SC0004410.

\end{document}